\documentclass[sigconf]{acmart}

\copyrightyear{2025}
\acmYear{2025}
\setcopyright{rightsretained}
\acmConference[HEART 2025]{The International Symposium on Highly Efficient Accelerators and Reconfigurable Technologies 2025}{May 26--28, 2025}{Kumamoto, Japan}
\acmBooktitle{The International Symposium on Highly Efficient Accelerators and Reconfigurable Technologies 2025 (HEART 2025), May 26--28, 2025, Kumamoto, Japan}
\acmDOI{10.1145/3728179.3728192}
\acmISBN{979-8-4007-1432-0/25/05}

\usepackage{tabularx}
\usepackage{graphicx}
\usepackage{subcaption}

\begin{document}

\title{ResBench: Benchmarking LLM-Generated FPGA Designs with Resource Awareness}

\author{Ce Guo}
\affiliation{
  \institution{Imperial College London}
  \city{}
  \country{United Kingdom}}
\email{c.guo@imperial.ac.uk}

\author{Tong Zhao}
\affiliation{
  \institution{Imperial College London}
  \city{}
  \country{United Kingdom}}
\email{tong.zhao24@imperial.ac.uk}

\renewcommand{\shortauthors}{C. Guo and T. Zhao}
\renewcommand{\shorttitle}{ResBench: A Resource-Aware Benchmark for LLM-Generated FPGA Designs}

\begin{abstract}
Field-Programmable Gate Arrays (FPGAs) are widely used in modern hardware design, yet writing Hardware Description Language (HDL) code for FPGA implementation remains a complex and time-consuming task. Large Language Models (LLMs) have emerged as a promising tool for HDL generation, but existing benchmarks for LLM-based code generation primarily focus on functional correctness while overlooking hardware resource usage. Furthermore, current benchmarks offer limited diversity and do not fully represent the wide range of real-world FPGA applications. To address these shortcomings, we introduce ResBench, the first resource-focused benchmark explicitly designed to distinguish between resource-optimized and inefficient LLM-generated HDL code. ResBench consists of 56 problems across 12 categories, covering applications from finite state machines to financial computing. Our open-source evaluation framework automatically tests LLMs by generating Verilog code, verifying correctness, and measuring resource usage. The experiments, which primarily analyze Lookup Table (LUT) usage, reveal significant differences among LLMs, demonstrating ResBench’s capability to identify models that generate more resource-optimized FPGA designs.

\end{abstract}


\maketitle

\section{Introduction}

Field-Programmable Gate Arrays (FPGAs) are widely used in reconfigurable computing, providing flexible and high-performance hardware implementations for applications such as artificial intelligence acceleration, financial computing, and embedded systems. However, FPGA development traditionally requires manual coding in hardware description languages (HDLs), a process that is both time-consuming and prone to errors. Large Language Models (LLMs) have recently shown potential in automating HDL generation, offering a way to improve FPGA design productivity.

While studies such as VeriGen~\cite{thakur2023benchmarking} and RTLLM~\cite{lu2024rtllm} have examined the feasibility of generating Verilog with LLMs, most existing benchmarks focus primarily on functional correctness while paying little attention to resource constraints, which is an essential consideration in FPGA design.

FPGA designs are subject to strict hardware resource constraints. Even when HDL code passes functional correctness tests, its efficiency in utilizing these resources can vary significantly depending on how logic is optimized and mapped to FPGA hardware. In reconfigurable computing, resource-aware optimizations play a crucial role in determining whether a design can be practically deployed and the degree of parallelism achievable. However, existing benchmarks for LLM-generated hardware designs typically evaluate the generated code based only on syntax and functional correctness, overlooking resource-related issues.

To address this limitation, we introduce \textbf{ResBench}, the first FPGA-resource-focused benchmark specifically designed to evaluate LLM-generated designs based on resource usage. Unlike previous benchmarks that focus primarily on syntax and functional correctness, ResBench highlights how well LLMs generate Verilog code optimized for FPGA resource utilization. Our key contributions are:

\begin{itemize}
\item A resource-focused benchmark featuring 56 problems across 12 categories, covering real-world FPGA workloads such as combinational logic, state machines, AI accelerators, and financial computing applications. (Section~\ref{sec:bench})
\item An open-source automated evaluation framework that performs LLM querying, functional correctness testing, FPGA synthesis, and resource measurement\footnote{Code repository: \url{https://github.com/jultrishyyy/ResBench}}. The framework automatically generates Verilog code using LLMs and evaluates its correctness and resource usage. (Section~\ref{sec:fw})
\item A detailed study of nine LLMs, comparing their performance in functional correctness and FPGA resource usage. The results reveal substantial differences in how various models generate resource-conscious designs. (Section~\ref{sec:eval})
\end{itemize}

By integrating FPGA resource awareness into benchmarking, ResBench provides a practical evaluation of LLM-generated HDL. This benchmark establishes a foundation for advancing AI-driven FPGA design, encouraging the development of more resource-efficient models optimized for FPGAs.

\section{Background}

This section explores the evolution of large language models (LLMs) from general code generation to hardware design generation using hardware description languages (HDLs). We examine their capabilities and limitations, as well as existing benchmarks for LLM-generated hardware design.

\begin{table*}[ht]
    \centering
    \caption{Benchmarks for LLM HDL Generation}
    \label{verilog_benchmark}
    \small
    \begin{tabular}{l c c p{3cm} p{7cm}}
    \toprule
    \textbf{Benchmark} & \textbf{Size} & \textbf{PL} & \textbf{Type} & \textbf{Features} \\
    \midrule
    \textbf{VerilogEval \cite{liu2023verilogeval}} & 156  & Verilog & Verilog code generation tasks & Covers a wide range of tasks from simple combinational circuits to finite state machines; includes automatic functional correctness testing. \\

    \midrule
    \textbf{HDLEval \cite{kashanaki2024hdleval}} & 100 & Multiple & Language-agnostic HDL evaluation & Evaluates LLMs across multiple HDLs using standardized testbenches and formal verification; categorizes problems into combinational and pipelined tests. \\

    \midrule
    \textbf{PyHDL-Eval \cite{batten2024pyhdl}} & 168  & Python-embedded DSLs & Specification-to-RTL tasks & Focuses on Python-embedded DSLs for hardware design; includes Verilog reference solutions and testbenches; evaluates LLMs' ability to handle specification-to-RTL translations. \\

    \midrule
    \textbf{RTLLM \cite{liu2024openllm}} & 50 & Verilog, VHDL, Chisel & Design RTL generation & Supports evaluation across multiple HDL formats; spans various design complexities and scales; includes an automated evaluation framework. \\

    \midrule
    \textbf{VHDL-Eval \cite{vijayaraghavan2024vhdl}} & 202  & VHDL & VHDL code generation tasks & Aggregates translated Verilog problems and publicly available VHDL problems; utilizes self-verifying testbenches for functional correctness validation. \\

    \midrule
    \textbf{GenBen \cite{wangenben}} & 351 & Verilog & Fundamental hardware design and debugging tasks & Evaluates synthesizability, power consumption, area utilization, and timing performance to ensure real-world applicability. \\
    \bottomrule
    \end{tabular}
\end{table*}

\subsection{Code-specialized and HDL-specialized LLMs}

Language models have seen significant advancements, particularly with the introduction of Transformers \cite{vaswani2017attention}. Large-scale pre-trained models such as BERT \cite{devlin2018bert}, the GPT series \cite{radford2018improving, brown2020language, ye2023comprehensive}, and PaLM 2 \cite{chowdhery2023palm} have expanded their capabilities across various tasks, including code generation.

Code-specialized LLMs are models designed to generate computer programs based on human-language prompts. Surveys such as \cite{jiang2024survey, zheng2023survey, wang2023review} review the latest techniques and model architectures developed for this purpose. In general, a code-specialized LLM can be created using two different approaches.
\begin{itemize}
\item The first approach trains models from scratch on large-scale open-source code datasets spanning multiple programming languages, as demonstrated by CodeGen \cite{nijkamp2022codegen}, InCoder \cite{fried2022incoder}, and StarCoder \cite{li2023starcoder}. These code-first LLMs excel at code completion and multi-language generation. However, this method demands extensive computational resources and high-quality datasets. Additionally, these models may struggle with instruction-driven tasks due to their limited ability to process human language instructions.
\item The second approach fine-tunes general-purpose LLMs for coding tasks, as seen in Codex \cite{chen2021evaluating}, which is derived from GPT-3, and Code Llama \cite{roziere2023code}, which builds on Meta’s Llama. Fine-tuning incorporates the linguistic knowledge of general LLMs, allowing them to handle natural language prompts while maintaining strong coding capabilities. This method is more computationally efficient than training from scratch, but it may lack the precision of code-first models in understanding programming language syntax.
\end{itemize}

While significant research has explored the use of language models for general software code generation, the application of LLMs to HDL code has not received comparable attention. Existing work on HDL code generation primarily focuses on improving general-purpose or software-code-specialized LLMs. In particular, these studies aim to improve these LLMs' understanding of hardware description tasks by training them on HDL datasets and benchmarking frameworks. Notable efforts in this area include VeriGen, MEV-LLM, and AutoVCoder.
\begin{itemize}
\item VeriGen \cite{thakur2024verigen} fine-tunes CodeGen (2B, 6B, 16B) on a Verilog dataset collected from GitHub repositories and textbooks. It employs supervised fine-tuning with problem-code pairs, validating functional correctness using a benchmark problem set and problems from HDLBits tutorials \cite{hdlbits2017}.
\item MEV-LLM \cite{nadimi2024multi} trains on 31,104 source code files from GitHub, labeled by GPT-3.5, to fine-tune CodeGen (2B, 6B, 16B) and GEMMA (2B, 7B) models. This approach yields an improvement of up to 23.9\% in the Pass@k metric \cite{chen2021evaluating} over VeriGen.
\item AutoVCoder \cite{gao2024autovcoder} also uses Verilog code from GitHub, filtering high-quality samples with ChatGPT-3.5. It applies a two-stage fine-tuning process to improve generalization, with final evaluation conducted on a real-world benchmark.
\end{itemize}

Beyond direct fine-tuning, reinforcement learning approaches like Golden Code Feedback is used to refine models iteratively using user feedback  \cite{wang2024large}. Similarly, multi-modal techniques such as VGV \cite{wong2024vgv} integrate circuit diagrams with textual data during training, allowing models to understand spatial and parallel aspects of circuit design.

\subsection{Benchmarks for LLM-Generated Software}

The research community has recognized the need for standardized benchmarks to rigorously evaluate LLM-generated code in terms of design correctness.

Most existing benchmarks for code generation are tailored for software development rather than hardware design. For instance, HumanEval \cite{chen2021evaluating} and MBPP (Mostly Basic Python Problems) \cite{liu2024your} are widely used to evaluate LLMs for software code generation. HumanEval consists of 164 Python programming tasks, each with a function signature and a set of test cases to validate correctness. While relatively small in scope, this benchmark is carefully designed, making it well-suited for quick functional correctness evaluations of Python-based code generation. Several extensions have expanded its coverage, including HumanEval+ \cite{liu2024your}, which increases the number of test cases by 80 times, and HumanEvalPack \cite{muennighoff2023octopack}, which extends the benchmark to six programming languages.

Similarly, MBPP comprises approximately 974 short Python programming tasks, each including a task description prompt, a code solution, and three automated test cases. It emphasizes both correctness and clarity. Its enhanced version, MBPP+ \cite{liu2024your}, refines flawed implementations and expands the number of test cases to improve robustness.

While HumanEval and MBPP provide fundamental benchmarks, they primarily focus on entry-level programming tasks and do not always reflect the complexity of real-world software development. To address this limitation, benchmarks with more intricate problem sets have been introduced. For example, DSP-1000 \cite{lai2023ds} contains 1,000 science-related programming tasks from seven Python libraries, covering a diverse set of topics and incorporating multi-criteria evaluation metrics to provide a more realistic assessment of code generation models.

\subsection{Benchmarks for LLM-Generated Hardware}

Existing benchmarks for hardware design generation in HDL are often based on experiences and lessons learned in software code benchmarking, particularly in designing diverse problem sets and developing automated evaluation frameworks. Similar to their software counterparts, performance metrics for hardware design generation frequently focus on design correctness measures such as Pass@k. However, benchmarking hardware designs typically involves simulation-based hardware verification \cite{qiu2024autobench}. 

Unlike software benchmarks, which evaluate correctness by directly executing code, hardware designs require dedicated testbench scripts to simulate hardware behavior and validate functionality. A typical evaluation process involves querying the LLM for HDL code, executing the generated HDL within a hardware simulation environment, and comparing the outputs to determine correctness.

Some benchmarks have emerged to address these challenges, as presented in Table~\ref{verilog_benchmark}. Among them, VerilogEval, HDLEval, and PyHDL-Eval are widely recognized.

\begin{itemize}
    \item VerilogEval \cite{liu2023verilogeval} is a widely adopted benchmark for evaluating LLMs in Verilog code generation \cite{abdelatty2024metrex,gao2024autovcoder, yang2025haven}. It consists of 156 problems taken from the HDLBits tutorial website \cite{hdlbits2017}, covering a range of Verilog tasks from combinational circuits to finite state machines. The framework automates functional correctness testing by comparing the simulation outputs of generated designs against predefined golden solutions.

    \item HDLEval \cite{kashanaki2024hdleval} follows a language-agnostic approach. In particular, this benchmark allows the same set of problems, formulated in plain English, to be evaluated across different HDLs. The benchmark consists of 100 problems systematically categorized into combinational and pipelined designs, covering fundamental hardware components such as logic gates, arithmetic operations, and pipelined processing units. A prominent feature of this benchmark is the use of formal verification instead of unit tests. This feature ensures that the generated HDL code is functionally correct and maintains logical equivalence with reference implementations.

    \item PyHDL-Eval \cite{batten2024pyhdl} is a framework for evaluating LLMs on specification-to-Register Transfer Level (RTL) tasks within Python-embedded domain-specific languages (DSLs). It includes 168 problems across 19 subclasses, covering combinational logic and sequential logic. The evaluation process involves executing the generated code in Python-embedded HDLs (e.g., PyMTL3, PyRTL) and measuring functional correctness based on pass rates.
\end{itemize}

\subsection{Challenges in LLM Benchmarking for FPGA Design}

Despite advancements in LLM-driven Verilog generation, existing models primarily focus on producing syntactically and functionally correct HDL but fail to address critical hardware constraints essential for FPGA deployment, such as resource efficiency, timing constraints, and power consumption. Unlike ASIC design, FPGA-based development demands careful consideration of resource usage, including lookup tables (LUTs), flip-flops (FFs), block RAM (BRAM), and digital signal processing (DSP) blocks. However, current LLMs for Verilog generation lack an understanding of FPGA-specific requirements, often producing designs that are functionally correct but inefficient and impractical for real-world FPGA deployment.

The inability to evaluate and optimize LLM-generated Verilog for FPGA resource constraints highlights the need for advancing resource-aware Verilog generation and motivates this study. To establish LLMs as a practical solution for HDL automation, it is essential to equip them with a deeper understanding of FPGA design constraints. Achieving this requires developing new training datasets, designing robust evaluation frameworks, and refining LLM training strategies to enhance their capability in hardware-aware code generation. This paper focuses on constructing a benchmark that provides an evaluation of LLMs' performance in generating HDLs that are both functionally correct and optimized for FPGA-specific resource constraints.

\section{Design of ResBench}
\label{sec:bench}
This section presents the design of ResBench, outlining its guiding principles and structured problem set for evaluating LLM-generated Verilog code. Additionally, we compare ResBench with existing benchmarks for LLM-generated HDL code.

\subsection{Design Principles and Benchmark Problems}
ResBench is designed to evaluate LLM-generated Verilog across a diverse range of FPGA applications, with a primary focus on resource optimization awareness.  The benchmark consists of 56 problems categorized into 12 domains, each representing a key area of FPGA applications. The problems in ResBench are carefully structured to evaluate both functional correctness and resource efficiency.  The benchmark covers a wide range of FPGA design tasks, from fundamental digital logic and mathematical computation to more complex, application-driven domains such as machine learning, cryptography, and financial computing. By spanning these diverse categories, ResBench ensures that LLMs are tested on both low-level design problems and high-level algorithmic implementations for real-world applications.

The design of ResBench is guided by two key principles:
\begin{itemize}
\item \textbf{The Principle of Resource Usage Differentiation} aims to highlight differences in how LLMs optimize FPGA resource usage. The benchmark includes problems that allow multiple resource-aware optimization strategies and require mathematical transformations. This approach makes it possible to distinguish between models that generate resource-efficient Verilog and those that do not.
\item \textbf{The Principle of FPGA Application Diversity} recognizes the wide range of FPGA applications, particularly in computational acceleration and edge computing. ResBench spans various domains, including financial computing, climate modeling, and signal processing, allowing LLMs to be tested across a broad set of real-world FPGA workloads.
\end{itemize}

\begin{table*}[t]
    \centering
    \caption{Summary of benchmark categories, including the number of problems and representative examples.}
    \label{tab:benchmark_summary}
    \begin{tabular}{l c l}
        \toprule
        \textbf{Category} & \textbf{\# Problems} & \textbf{Example Problems} \\
        \midrule
        Combinational Logic & 8 & parity\_8bit, mux4to1, bin\_to\_gray \\
        Finite State Machines & 4 & fsm\_3state, traffic\_light, elevator\_controller \\
        Mathematical Functions & 5 & int\_sqrt, fibonacci, mod\_exp \\
        Basic Arithmetic Operations & 5 & add\_8bit, mult\_4bit, abs\_diff \\
        Bitwise and Logical Operations & 4 & bitwise\_ops, left\_shift, rotate\_left \\
        Pipelining & 5 & pipelined\_adder, pipelined\_multiplier, pipelined\_fir \\
        Polynomial Evaluation & 5 & $(x+2)^2 + (x+2)^2 + (x+2)^2$, $(a+b)^2 - (a-b)^2$ \\
        Machine Learning & 5 & matrix\_vector\_mult, relu, mse\_loss \\
        Financial Computing & 4 & compound\_interest, present\_value, currency\_converter \\
        Encryption & 3 & caesar\_cipher, modular\_add\_cipher, feistel\_cipher \\
        Physics & 4 & free\_fall\_distance, kinetic\_energy, wavelength \\
        Climate & 4 & carbon\_footprint, heat\_index, air\_quality\_index \\
        \midrule
        \textbf{Total} & \textbf{56} & -- \\
        \bottomrule
    \end{tabular}
\end{table*}

\begin{figure*}[t]
    \centering
    \begin{subfigure}{0.4\textwidth}
        \centering
        \includegraphics[width=\linewidth]{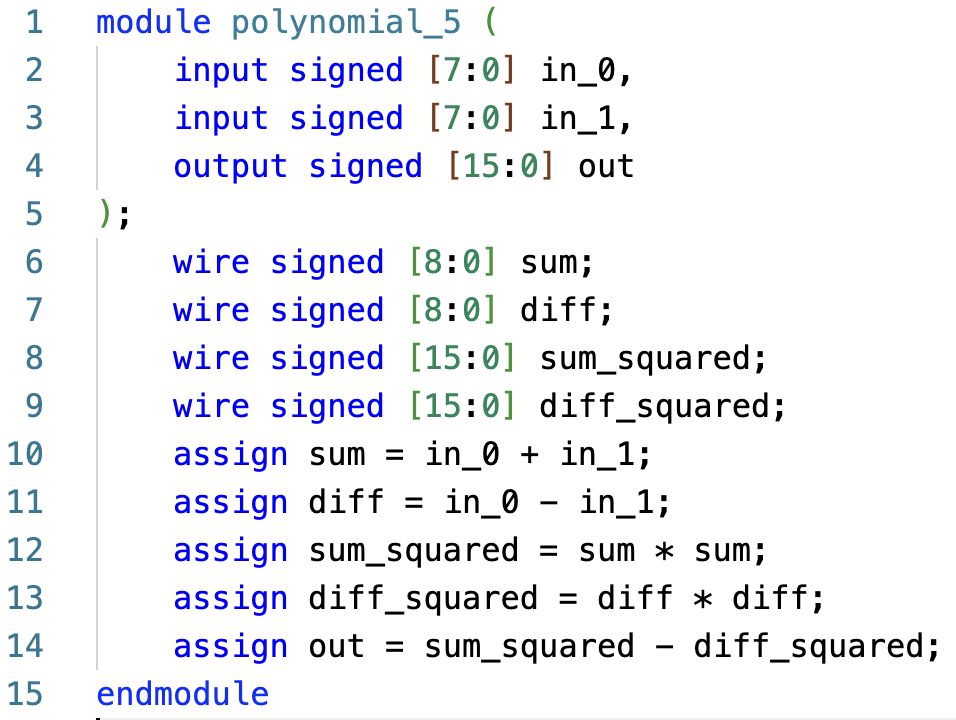}
        \caption{Design Generated by Qwen-2.5 (213 LUTs)}
        \label{fig:poly_qwen}
    \end{subfigure}
    \hfill
    \begin{subfigure}{0.45\textwidth}
        \centering
        \includegraphics[width=\linewidth]{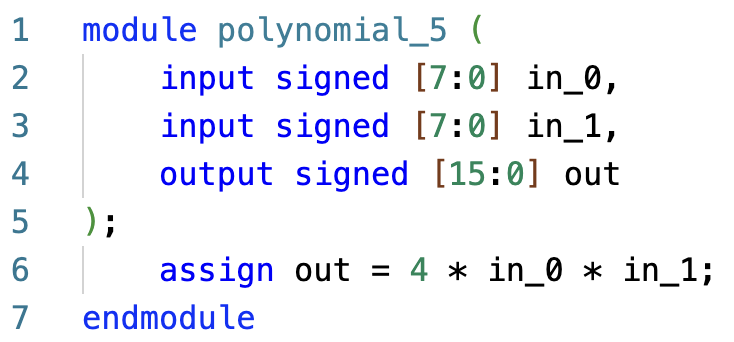}
        \caption{Design Generated by GPT-4 (0 LUT + 1 DSP)}
        \label{fig:poly_gpt4}
    \end{subfigure}
   \caption{Benchmark example illustrating HDL optimization capability using the expression \((a+b)^2 - (a-b)^2\). 
(a) Qwen-2.5 computes the full expression directly, leading to high LUT usage. 
(b) GPT-4 simplifies the expression to \(4ab\), significantly reducing resource usage by using a single DSP unit instead of LUTs. This example demonstrates ResBench's ability to differentiate LLMs based on resource optimization.}
    \label{fig:poly_comparison}
    \Description{Benchmark example illustrating HDL optimization capability for minimizing LUT usage using the expression \((a+b)^2 - (a-b)^2\). Qwen-2.5 computes the full expression directly, leading to high LUT usage.  GPT-4 simplifies the expression to \(4ab\), significantly reducing resource usage by using a single DSP unit instead of LUTs.}
\end{figure*}

Table~\ref{tab:benchmark_summary} provides an overview of the problem categories along with representative examples. The problems in the benchmark are designed to align with the two guiding principles. Specifically, ResBench addresses the principles as follows:
\begin{itemize}
    \item ResBench addresses resource usage differentiation by introducing problems that require optimization techniques beyond simple code-level improvements. Fig.~\ref{fig:poly_comparison} illustrates this with a polynomial evaluation problem from the benchmark. Algebraic simplification in this case enables a more efficient Verilog implementation using fewer LUTs. LLMs that fail to apply this optimization generate designs with excessive LUT consumption.

    \item ResBench addresses FPGA application diversity by covering both foundational and application-driven workloads. Foundational problems include combinational logic, finite state machines, arithmetic operations, pipelining, polynomial evaluations, and mathematical functions. Application-driven problems span machine learning, encryption, and financial computing, emphasizing the role of FPGAs in AI acceleration, security, and high-speed data processing.
\end{itemize}

By adhering to these principles, the benchmark evaluates not only an LLM's ability to generate syntactically correct Verilog code but also its capability to produce hardware-efficient designs suited for FPGA deployment.

\subsection{Comparison with Existing Benchmarks}

\begin{table*}[t]
    \centering
    \caption{Comparison of Benchmarks for LLM HDL Evaluation}
    \begin{tabular}{lcccc}
        \toprule
        \textbf{Benchmark} & \textbf{Year} & \textbf{Hardware} & \textbf{Optimization} & \textbf{Problem Diversity} \\
        &&&\textbf{Awareness}&\\
        \midrule
        VerilogEval \cite{thakur2024verigen} & 2023 & General HDL & No & Logic, FSMs, arithmetic \\
        HDLEval \cite{kashanaki2024hdleval} & 2024 & General HDL & No & Digital circuits, control logic \\
        PyHDL-Eval \cite{batten2024pyhdl} & 2024 & General HDL & No & Python-based HDL, small designs \\
        RTLLM \cite{lu2024rtllm} & 2024 & General HDL & No & RTL, bus protocols, DSP \\
        VHDL-Eval \cite{vijayaraghavan2024vhdl} & 2024 & General HDL & No & VHDL logic, sequential circuits \\
        GenBen \cite{wangenben} & 2024 & General HDL & No & Application-driven tasks \\
        \midrule
        ResBench (This paper) & 2025 & FPGA & Resource Usage Optimizations & 56 problems across 12 domains \\
        \bottomrule
    \end{tabular}
    \label{tab:benchmark_comparison}
\end{table*}

Table~\ref{tab:benchmark_comparison} highlights key differences between ResBench and existing HDL benchmarks. The differences are particularly significant in FPGA resource optimization awareness and problem diversity:
\begin{itemize}
\item FPGA resource optimization awareness. Most existing benchmarks, such as VerilogEval, HDLEval, and GenBen, focus primarily on functional correctness and HDL syntax quality but do not explicitly account for FPGA resource usage. Consequently, these benchmarks cannot distinguish between functionally correct designs that differ significantly in hardware resource utilization. In contrast, ResBench introduces optimization-aware problems specifically designed to expose variations in resource usage. This enables a more practical comparison of LLMs based on their ability to generate resource-efficient designs for FPGAs.

\item Problem diversity. Existing benchmarks primarily focus on fundamental HDL constructs such as basic logic, state machines, and arithmetic operations, with limited diversity in FPGA applications. For example, VerilogEval emphasizes control logic and arithmetic, while HDLEval mainly evaluates digital circuits and state machines. In contrast, ResBench encompasses a significantly broader range of FPGA applications, including machine learning, encryption, financial computing, and physics-based modeling. These domains represent real-world FPGA workloads where resource efficiency is crucial for minimizing device cost and maximizing parallelism. By incorporating a diverse set of tasks, ResBench provides a more comprehensive evaluation of LLM-generated HDL in practical FPGA design scenarios.
\end{itemize}

\section{Evaluation Framework for ResBench}
\label{sec:fw}

\begin{figure*}[t]
    \centering
    \includegraphics[width=0.95\textwidth]{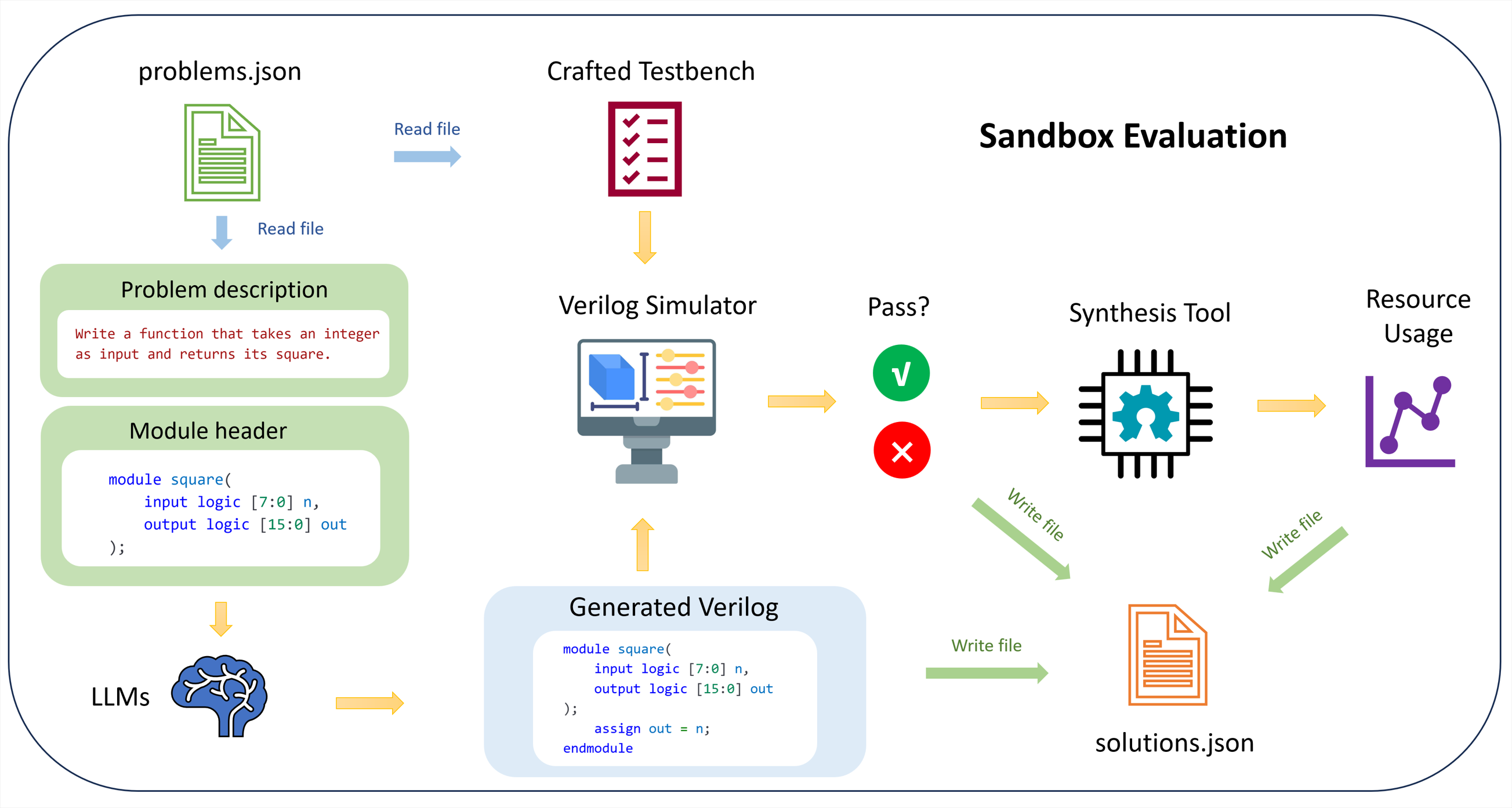} 
    \Description{A high-level overview of the evaluation pipeline, consisting of Verilog generation, functional verification, FPGA synthesis, and report generation.}
    \caption{Overview of the software workflow. The process begins with Verilog generation using an LLM, followed by functional verification through testbenches. Functionally correct designs undergo FPGA synthesis to extract resource usage metrics, and the framework compiles performance reports comparing functional correctness and resource usage.}
    \label{evaluation} 
\end{figure*}

To evaluate LLM-generated designs for FPGA design with ResBench, we implement a structured framework that examines both functional correctness and hardware efficiency.

We build the software for the evaluation framework based on the lessons learned in testing LLM-based software code generation. The benchmarks for LLM-based software share a common evaluation framework aimed at assessing whether generated code is both syntactically valid and functionally correct. In particular, each benchmark generally includes four key components:
\begin{enumerate}
\item Prompts, which can be presented as a natural language description \cite{hendrycks2021measuring, uniyal2024one} or both description and function signature \cite{chen2021evaluating, liu2024your}, guiding the model on what to generate.
\item A reference solution, which serves as the correct implementation for comparison.
\item Test cases, which are predefined inputs and expected outputs used to validate correctness
\item Performance metrics, like Pass@k \cite{chen2021evaluating} and Code Similarity Scores \cite{ren2020codebleu}, which estimates how effectively an LLM-generated solution satisfies the given problem constraints. A widely used metric is Pass@k, which measures the likelihood that at least one of the top-k generated solutions passes all test cases.
\end{enumerate}

Different from the evaluation of software code generation, our framework uses the resource usage count as a key metric to quantify the quality of resource-oriented optimization. The framework also uses an automated benchmarking system that automates Verilog code generation, functional correctness testing, FPGA synthesis, and resource usage extraction.

The open-source software for ResBench automates the evaluation of LLM-generated Verilog, systematically measuring both functional correctness and FPGA resource usage with minimal manual intervention. An overview of the software's workflow is shown in Fig.~\ref{evaluation}.

The software accepts a user-specified LLM and generates Verilog solutions based on structured problem definitions. It produces detailed evaluation reports, indicating whether each design passes synthesis and functional correctness checks, along with a resource utilization summary. By automating the full evaluation pipeline, the framework facilitates large-scale benchmarking and comparative studies of different LLMs for HDL code generation.

To maintain consistency, each problem follows a structured format consisting of three components: a natural language problem description in plain English, a Verilog module header, and a predefined testbench. The problem description specifies the expected input-output format and functional constraints, ensuring that LLM-generated code aligns with real-world design requirements. The module header provides a consistent Verilog interface with defined input and output signals but leaves the internal logic for the LLM to fill. The testbench validates functional correctness through simulation by applying predefined test cases in the testbench and comparing outputs against a manually verified reference solution.

The evaluation framework follows a structured process to evaluate LLM-generated designs. The evaluation of an LLM on a benchmark problem consists of the following three steps:
\begin{enumerate}
\item The framework queries the selected LLM to generate multiple Verilog code snippets for a given problem. These generated snippets are stored in text format along with references to their corresponding problem descriptions. Functional correctness is then verified using predefined testbenches. Designs that pass all test cases proceed to FPGA synthesis, while those that fail have their errors recorded for further analysis.
\item FPGA synthesis is performed to determine resource usage metrics such as LUT count, DSP utilization, and register count. For designs that fail synthesis, the resource count is set to \( \infty \), ensuring a consistent comparison framework.
\item The framework generates structured reports summarizing pass rates, synthesis success rates, and resource usage statistics. Users can visualize model performance through automatically generated comparisons of functional correctness and resource usage.
\end{enumerate}
By following this structured evaluation process, the framework provides a fully automated benchmarking solution that evaluates LLM-generated Verilog across all benchmark problems, focusing on both design correctness and resource usage.

\section{Evaluation}
\label{sec:eval}
This section presents our evaluation of LLM-generated FPGA designs using ResBench. The evaluation focuses not only on functional correctness but also on FPGA resource usage.

\subsection{Experimental Setup and Metrics}

Our experiments evaluate the capability of LLMs to generate both funtionally correct and resource-efficient Verilog code by examining the number of functionally correct designs. Also, we examine how well different models optimize FPGA resources.

We run the software proposed in Section~\ref{sec:fw} with all benchmark problems. For each LLM-generated design, we compile the testbench and simulate it to verify the functional correctness of the generated Verilog modules. If the simulation confirms that the design is correct, we use the Vivado synthesis tool to generate a resource report and assess resource usage.

To evaluate functional correctness, we generate the same number of designs from each LLM and count how many designs pass for each problem. In cases where two LLMs produce the same number of passing designs, we break the tie by considering the number of designs that pass synthesis but fail the functional correctness test. We do not use the Pass@k metric, which is commonly applied to LLM-generated software code, because we intend to distinguish designs that fail hardware synthesis from those that successfully synthesize but do not meet functional correctness requirements.

In this study, we quantify the capability of resource optimization by minimizing the LUT count. LUTs serve as the primary logic resource for implementing combinational operations and small memory elements. While FPGAs also provide other resources, these tend to be application-specific. For instance, DSPs and BRAMs are crucial for arithmetic-intensive and memory-heavy designs but are not universally required across all FPGA applications. In contrast, LUTs are a fundamental component in nearly every design, making them a consistent and reliable metric for evaluating different HDL implementations.

Each LLM-generated design \( d_i \) is evaluated based on its LUT usage. If a design successfully passes both synthesis and functional correctness testing, its LUT count is recorded as \( \text{LUT}(d_i) \). Otherwise, it is assigned \( \infty \) to indicate that the design is either non-synthesizable or functionally incorrect:

\begin{align}
\text{LUT}(d_i) =
\begin{cases}
\text{LUT count}, & \text{if } d_i \text{ is synthesizable and correct} \\
\infty, & \text{otherwise}
\end{cases}
\label{equ:lutdi}
\end{align}

\begin{align}
\text{LUT}_{\min} = \min \left( \text{LUT}(d_0), \text{LUT}(d_1), \dots, \text{LUT}(d_{n-1}) \right)
\label{equ:minlut}
\end{align}

By using \( \infty \) for failed designs, our approach naturally excludes non-functional implementations. This setting maintains computational consistency and eliminates the need for explicit filtering in the evaluation of Equation~\ref{equ:minlut}. Note that while this study focuses on minimizing LUT usage, our framework is capable of extracting and analyzing other resource metrics with a different optimization objective.

For correctness testing and resource usage evaluation, we use Vivado 2023.1 for simulation, synthesis, and analysis. The hardware implementation is targeted at the programmable logic section of the AMD Zynq 7000 XC7Z020CLG400-1 SoC, operating at its default clock frequency. While design correctness remains independent of the chosen tool, the LUT count is influenced by Vivado’s synthesis capabilities and the type of LUTs on the target device. However, we expect the impact of FPGA software choice on relative resource efficiency to be small. In particular, for a given problem, the HDL designs with the smallest LUT count will likely stay unchanged even when evaluated with different FPGA tools.

For all the evaluated LLMs, we set the temperature parameter to 1.5 to encourage high diversity of the HDL code for each problem. The evaluation includes three types of models: general-purpose LLMs, code-specialized LLMs, and HDL-specialized LLMs. The general-purpose models we evaluate include GPT-3.5 \cite{ye2023comprehensive}, GPT-4o \cite{hurst2024gpt}, GPT-4 \cite{achiam2023gpt}, GPT-o1-mini \cite{jaech2024openai}, Llama3.1-450B \cite{touvron2023llama}, Qwen-Max \cite{bai2023qwen}, and Qwen-Plus \cite{yang2024qwen2}. The evaluated code-specialized models include Qwen2.5-Coder-32B-Instruct \cite{hui2024qwen2} and Codestral \cite{jiang2023mistral}. We also evaluate VeriGen \cite{thakur2024verigen}, an HDL-specialized model. However, during the evaluation, VeriGen failed to generate legitimate Verilog code for all problems. As a result, we omit its results from further discussion.

\subsection{Functional Correctness}
\label{sec:eval:fc}
\begin{table*}[t]
    \centering
    \caption{Design correctness of LLMs in generating Verilog code across different categories}
    \label{passcount}
    Cell format: pass / synthesis OK but incorrect design / synthesis error
    \renewcommand{\arraystretch}{1.0}
    \footnotesize
    \begin{tabular}{cccccccccc} 
        \toprule
         & \textbf{GPT-3.5} & \textbf{GPT-4} & \textbf{GPT-4o} & \textbf{GPT-o1} & \textbf{Llama3.1} & \textbf{Qwen-max} & \textbf{Qwen-plus} & \textbf{Qwen2.5-coder} & \textbf{Codestral} \\
            & \textbf{turbo} & & & \textbf{mini} & \textbf{405B} & & & \textbf{32B} &
            \\
        \toprule
        \textbf{Combinational Logic} & 112 / 5 / 3 & 117 / 3 / 0 & \textbf{120 / 0 / 0} & 118 / 1 / 1 & 115 / 2 / 3 & 117 / 2 / 1 & 109 / 1 / 10 & 112 / 2 / 6 & \textbf{120 / 0 / 0} \\
        \midrule
        \textbf{Finite State Machines} & 23 / 15 / 22 & 32 / 22 / 6 & 31 / 24 / 5 & \textbf{39 / 18 / 3} & 31 / 24 / 5 & 34 / 26 / 0 & 27 / 23 / 10 & 39 / 10 / 11 & 36 / 6 / 18 \\
        \midrule
        \textbf{Mathematical Functions} & 13 / 19 / 43 & 6 / 39 / 30 & 36 / 10 / 29 & \textbf{46 / 24 / 5} & 7 / 6 / 62 & 26 / 27 / 22 & 20 / 26 / 29 & 5 / 8 / 62 & 0 / 3 / 72 \\
        \midrule
        \textbf{Basic Arithmetic Ops} & 37 / 2 / 36 & 63 / 8 / 4 & 66 / 9 / 0 & \textbf{68 / 4 / 3} & 43 / 2 / 30 & 38 / 22 / 15 & 27 / 13 / 35 & 54 / 6 / 15 & 62 / 13 / 0 \\
        \midrule
        \textbf{Bitwise \& Logic Ops} & 35 / 0 / 25 & 55 / 0 / 5 & 58 / 2 / 0 & \textbf{59 / 0 / 1} & 52 / 0 / 8 & 47 / 0 / 13 & 33 / 11 / 16 & 36 / 0 / 24 & 55 / 0 / 5 \\
        \midrule
        \textbf{Pipelining} & 0 / 59 / 16 & 11 / 54 / 10 & 26 / 49 / 0 & 15 / 38 / 22 & 7 / 38 / 30 & 15 / 32 / 28 & 16 / 26 / 33 & \textbf{21 / 31 / 23} & 6 / 56 / 13 \\
        \midrule
        \textbf{Polynomial Evaluation} & 19 / 3 / 53 & 69 / 0 / 6 & \textbf{74 / 1 / 0} & 68 / 5 / 2 & 58 / 6 / 11 & 55 / 2 / 18 & 28 / 5 / 42 & 65 / 7 / 3 & 69 / 6 / 0 \\
        \midrule
        \textbf{Machine Learning} & 31 / 3 / 41 & 60 / 8 / 7 & 60 / 13 / 2 & \textbf{73 / 1 / 1} & 45 / 28 / 2 & 63 / 12 / 0 & 61 / 12 / 2 & 57 / 2 / 16 & 64 / 8 / 3 \\
        \midrule
        \textbf{Financial Computing} & 9 / 23 / 28 & 21 / 22 / 17 & \textbf{29 / 13 / 18} & 20 / 20 / 20 & 11 / 21 / 28 & 28 / 15 / 17 & 15 / 12 / 33 & 16 / 7 / 37 & 17 / 23 / 20 \\
        \midrule
        \textbf{Encryption} & 30 / 0 / 15 & \textbf{30 / 2 / 13} & 25 / 20 / 0 & 30 / 0 / 15 & 26 / 0 / 19 & 25 / 9 / 11 & 30 / 1 / 14 & 30 / 0 / 15 & 30 / 0 / 15 \\
        \midrule
        \textbf{Physics} & 45 / 3 / 12 & \textbf{57 / 0 / 3} & 53 / 4 / 3 & 54 / 5 / 1 & 41 / 11 / 8 & 49 / 7 / 4 & 40 / 17 / 3 & 38 / 15 / 7 & 55 / 2 / 3 \\
        \midrule
        \textbf{Climate} & 8 / 15 / 37 & 21 / 30 / 9 & \textbf{41 / 11 / 8} & 41 / 15 / 4 & 24 / 23 / 13 & 38 / 19 / 3 & 19 / 31 / 10 & 32 / 14 / 14 & 28 / 19 / 13 \\
        \bottomrule
        \textbf{Number of wins} & \textbf{0} & \textbf{2} & \textbf{4} & \textbf{5} & \textbf{0} & \textbf{0} & \textbf{0} & \textbf{1} & \textbf{1} \\
        \bottomrule
    \end{tabular}

\end{table*}

Table~\ref{passcount} provides detailed pass counts across 12 categories of problems, with 15 designs generated for each problem. Each table cell follows the format: \textbf{pass / synthesis OK but incorrect design / synthesis error}. For example, if a category contains 5 problems, each LLM generates a total of 75 solutions (5 problems × 15 generated designs per problem), and the sum of the three numbers in each cell corresponds to this total. In this table, we also include the number of wins, which represents the number of categories in which each LLM achieved the highest pass count.

The results show that GPT-o1-mini is the leading model, achieving the highest pass counts in most categories. This suggests that reasoning-optimized models have an advantage in Verilog code generation. This is potentially because its reasoning capabilities contribute to more accurate outputs.

Table~\ref{passcount} shows a notable observation in finite state machines, mathematical functions, pipelining. The generated code can often pass synthesis but fail to function correctly. This observation suggests that while LLMs grasp basic syntax, they struggle with complex functional logic. In contrast, for more intricate problems with complex contexts, such as mathematical functions and financial computing, LLMs tend to produce syntactically incorrect code, reflecting challenges in understanding and reasoning within these contexts.

The results show that for every problem there is at least one model providing correct solutions. However, in categories such as pipelining, financial computing, and encryption, LLMs tend to underperform and show higher variability. For example, GPT-3.5 produced no passing solutions in pipelining, but LLaMA 3.1 achieved good results. A similar pattern is observed in the mathematical functions category. These results highlight the importance of evaluating both functional correctness and resource optimization in complex design scenarios.

\subsection{Resource Usage}

\begin{table*}[t]
    \centering
    \caption{$\text{LUT}_{\min}$ for each LLM across categories}
    {
    \small
    \renewcommand{\arraystretch}{1.0}
    \begin{tabular}{cccccccccc}
            \toprule
             & \textbf{GPT-3.5} & \textbf{GPT-4} & \textbf{GPT-4o} & \textbf{GPT-o1} & \textbf{Llama3.1} & \textbf{Qwen-max} & \textbf{Qwen-plus} & \textbf{Qwen2.5-coder} & \textbf{Codestral} \\
            & \textbf{turbo} & & & \textbf{mini} & \textbf{405B} & & & \textbf{32B} &
            \\
            \toprule
            \textbf{fsm 3state} & 1 & \textbf{0} & \textbf{0} & \textbf{0} & \textbf{0} & \textbf{0} & \textbf{0} & \textbf{0} & \textbf{0} \\
            \midrule
            \textbf{traffic light} & 1 & 1 & 2 & \textbf{0} & \textbf{0} & 2 & 3 & 2 & $\infty$ \\
            \midrule
            \textbf{elevator controller} & 3 & 3 & \textbf{2} & \textbf{2} & \textbf{2} & \textbf{2} & \textbf{2} & \textbf{2} & \textbf{2} \\
            \midrule
            \textbf{vending machine} & \textbf{1} & \textbf{1} & 2 & \textbf{1} & 2 & \textbf{1} & \textbf{1} & 2 & \textbf{1} \\
            \midrule
            \textbf{int sqrt} & $\infty$ & $\infty$ & 68 & 177 & $\infty$ & \textbf{64} & 229 & 173 & $\infty$ \\
            \midrule
            \textbf{fibonacci} & $\infty$ & 56 & \textbf{1} & 56 & 56 & 56 & $\infty$ & $\infty$ & $\infty$ \\
            \midrule
            \textbf{mod exp} & $\infty$ & $\infty$ & 4466 & 4669 & $\infty$ & 1911 & \textbf{1678} & $\infty$ & $\infty$ \\
            \midrule
            \textbf{power} & $\infty$ & \textbf{79} & 93 & 93 & $\infty$ & 93 & 93 & 93 & $\infty$ \\
            \midrule
            \textbf{log2 int} & $\infty$ & $\infty$ & $\infty$ & \textbf{10} & 20 & $\infty$ & $\infty$ & 12 & $\infty$ \\
            \midrule
            \textbf{abs diff} & \textbf{12} & \textbf{12} & 14 & \textbf{12} & \textbf{12} & $\infty$ & \textbf{12} & \textbf{12} & \textbf{12} \\
            \midrule
            \textbf{modulo op} & \textbf{82} & \textbf{82} & \textbf{82} & \textbf{82} & 111 & $\infty$ & $\infty$ & $\infty$ & $\infty$ \\
            \midrule
            \textbf{left shift} & \textbf{10} & \textbf{10} & \textbf{10} & \textbf{10} & \textbf{10} & 12 & 12 & \textbf{10} & \textbf{10} \\
            \midrule
            \textbf{pipelined adder} & $\infty$ & \textbf{0} & 16 & $\infty$ & \textbf{0} & $\infty$ & \textbf{0} & 15 & $\infty$ \\
            \midrule
            \textbf{pipelined multiplier} & $\infty$ & $\infty$ & 77 & 70 & \textbf{56} & $\infty$ & 70 & $\infty$ & $\infty$ \\
            \midrule
            \textbf{pipelined max finder} & $\infty$ & \textbf{0} & 24 & \textbf{0} & 24 & 24 & 24 & 24 & 24 \\
            \midrule
            \textbf{$x^3 + 3x^2 + 3x + 1$} & 49 & 49 & \textbf{0} & 91 & \textbf{0} & 91 & \textbf{0} & 91 & 49 \\
            \midrule
            \textbf{$(x+2)^2 + (x+2)^2 + (x+2)^2$} & 64 & 33 & 96 & \textbf{11} & 108 & 108 & 26 & 18 & 33 \\
            \midrule
            \textbf{$(a+b)^2 - (a-b)^2$} & $\infty$ & \textbf{0} & 213 & 59 & 16 & 213 & 16 & 16 & 16 \\
            \midrule
            \textbf{relu} & \textbf{8} & \textbf{8} & \textbf{8} & \textbf{8} & \textbf{8} & 16 & \textbf{8} & \textbf{8} & 16 \\
            \midrule
            \textbf{mse loss} & $\infty$ & 216 & \textbf{64} & \textbf{64} & 216 & \textbf{64} & 216 & \textbf{64} & \textbf{64} \\
            \midrule
            \textbf{compound interest} & $\infty$ & 13060 & 10135 & 10135 & 52950 & \textbf{9247} & $\infty$ & 10135 & 52950 \\
            \midrule
            \textbf{currency converter} & $\infty$ & $\infty$ & \textbf{0} & \textbf{0} & 25 & \textbf{0} & $\infty$ & $\infty$ & $\infty$ \\
            \midrule
            \textbf{free fall distance} & \textbf{6} & \textbf{6} & 64 & \textbf{6} & \textbf{6} & 64 & 67 & 64 & \textbf{6} \\
            \midrule
            \textbf{kinetic energy} & 70 & 70 & \textbf{54} & \textbf{54} & \textbf{54} & \textbf{54} & \textbf{54} & \textbf{54} & \textbf{54} \\
            \midrule
            \textbf{potential energy} & 6 & 6 & 84 & \textbf{0} & 6 & 6 & 6 & 6 & 6 \\
            \midrule
            \textbf{carbon footprint} & 174 & 121 & 110 & \textbf{92} & 121 & 121 & 110 & 110 & 110 \\
            \midrule
            \textbf{heat index} & \textbf{16} & \textbf{16} & 201 & \textbf{16} & 195 & \textbf{16} & 124 & 201 & 201 \\
            \midrule
            \textbf{air quality index} & $\infty$ & $\infty$ & 128 & \textbf{104} & $\infty$ & \textbf{104} & 116 & 128 & 128 \\
            \bottomrule
            \textbf{Number of wins} & \textbf{7} & \textbf{12} & \textbf{10} & \textbf{19} & \textbf{11} & \textbf{10} & \textbf{9} & \textbf{7} & \textbf{8} \\
            \bottomrule
        \end{tabular}
    }
    \label{tab:llm_lut}
\end{table*}

Table~\ref{tab:llm_lut} presents the LUT usage results for the benchmark problems. In this table, problems where all LLMs yield identical LUT usage are excluded for brevity. The cell with the lowest resource usage in each category is highlighted in bold. The number of wins is determined by counting these highlighted cells.

The results suggest that different LLMs have varying levels of optimization capability within our benchmark framework. This highlights how the benchmark problems reveal differences in LLMs' ability to optimize resource usage. Moreover, we have the following observations based on the results:
\begin{enumerate}
    \item GPT-o1-mini leads with 19 wins, significantly outperforming the runner-up, GPT-4, which achieves 12 wins. This indicates that GPT-o1-mini generates Verilog designs with lower resource usage in most problems. Its strong performance suggests that advanced reasoning capabilities may enhance its understanding of problem requirements, the Verilog language, and the complexities of hardware design. In contrast, code-specialized LLMs, while demonstrating high accuracy in producing functionally correct Verilog, may lack the reasoning depth needed to optimize designs effectively for FPGA constraints. This difference highlights that generating syntactically correct HDL alone is insufficient for producing resource-efficient hardware, as true optimization demands a deeper understanding of both design requirements and FPGA-specific constraints.
    \item GPT-3.5-Turbo, Qwen2.5-Coder, and Codestral demonstrate the weakest resource optimization ability, achieving only 7, 7, and 8 wins, respectively. The poor resource optimization of GPT-3.5-turbo is potentially due to its model size and lack of updates. Qwen2.5-Coder and Codestral, the two code-specialized models in our evaluation, also struggle with resource optimization. One possible explanation is that these models are primarily trained on software code rather than HDL, which may limit their ability to account for FPGA resource constraints when generating and optimizing Verilog. Additionally, key optimization techniques, such as mathematical simplifications, are unlikely to be picked up effectively from software code data.
    \item The observed variations in resource usage across different problem types confirm that our benchmarks can lead to divergent hardware resource usage for Verilog designs generated by different LLMs. For simple tasks such as combinational logic and basic arithmetic operations, the differences tend to be less significant. This is likely because the training data of the models include well-established reference solution for these problems. However, for more complex problems, our benchmark problems lead to significantly greater divergence in LUT usage. This suggests that our benchmark problems effectively evaluate the ability of LLMs to optimize resource usage beyond learned patterns.
\end{enumerate}

Considering both functional correctness and resource usage, we find that GPT-o1-mini achieves the highest performance in both aspects, while code-specialized models, including Qwen2.5-Coder and Codestral, perform the worst.

\section{Conclusion and Future Work}

LLMs provide a promising solution for automating HDL generation. However, most current benchmarks focus mainly on functional correctness while overlooking FPGA resource constraints. This lack of attention on FPGA resource efficiency underscores the need for resource-aware benchmarks to better evaluate LLM-generated HDL for real-world FPGA deployment. Additionally, current benchmarks lack problem diversity, limiting their effectiveness in evaluating real-world FPGA applications. To address these limitations, we introduce ResBench, the first resource-centric benchmark for LLM-generated HDL. ResBench features 56 problems spanning 12 categories. The benchmark problems are designed to expose LLMs' ability to generate Verilog designs optimized for FPGA resource usage. 

While ResBench is not explicitly designed to emphasize combinational logic and arithmetic operations, the current problem set naturally includes a high proportion of such designs. Future work will expand the benchmark to include more sequential designs, such as pipelined architectures and state-driven circuits. Additionally, although the current evaluation focuses on Verilog, our framework is designed to support multiple HDLs. Future efforts will extend support to VHDL and high-level synthesis (HLS) tools.

\begin{acks}
We thank the anonymous reviewers for their valuable feedback and suggestions. The support of the United Kingdom EPSRC (grant number UKRI256, EP/V028251/1, EP/N031768/1, EP/S030069/1, and EP/X036006/1), Intel, and AMD is gratefully acknowledged.

\textbf{AI Usage Statement:}
This work involves the use of generative AI in multiple aspects. The methodology presented in this paper focuses on evaluating the ability of AI models to generate HDL code. As such, all experimental results are based on Verilog designs produced by LLMs. For writing, ChatGPT-4 and Llama 3.1 were used to refine phrasing, improve clarity, and proofread the text.
\end{acks}

\clearpage
% \printbibliography
\bibliographystyle{ACM-Reference-Format}
\bibliography{hwllm}

@article{vaswani2017attention,
  title = {Attention is All You Need},
  author = {Vaswani, Ashish},
  journal = {Advances in Neural Information Processing Systems},
  year = {2017}
}

@article{thakur2024verigen,
  title = {{VeriGen}: A Large Language Model for {Verilog} Code Generation},
  author = {Thakur, Shailja and Ahmad, Baleegh and Pearce, Hammond and Tan, Benjamin and Dolan-Gavitt, Brendan and Karri, Ramesh and Garg, Siddharth},
  journal = {ACM Transactions on Design Automation of Electronic Systems},
  volume = {29},
  number = {3},
  pages = {1--31},
  year = {2024},
  publisher = {ACM New York, NY}
}

@article{brown2020language,
  title = {Language Models are Few-Shot Learners},
  author = {Brown, Tom and Mann, Benjamin and Ryder, Nick and Subbiah, Melanie and Kaplan, Jared D and Dhariwal, Prafulla and Neelakantan, Arvind and Shyam, Pranav and Sastry, Girish and Askell, Amanda and others},
  journal = {Advances in Neural Information Processing Systems},
  volume = {33},
  pages = {1877--1901},
  year = {2020}
}

@misc{devlin2018bert,
  title = {{BERT}: Pre-Training of Deep Bidirectional Transformers for Language Understanding},
  author = {Devlin, Jacob},
  journal = {arXiv preprint arXiv:1810.04805},
  year = {2018}
}

@misc{radford2018improving,
  title = {Improving Language Understanding by Generative Pre-Training},
  author = {Radford, Alec},
  year = {2018}
}

@misc{ye2023comprehensive,
  title = {A Comprehensive Capability Analysis of {GPT-3} and {GPT-3.5} Series Models},
  author = {Ye, Junjie and Chen, Xuanting and Xu, Nuo and Zu, Can and Shao, Zekai and Liu, Shichun and Cui, Yuhan and Zhou, Zeyang and Gong, Chao and Shen, Yang and others},
  journal = {arXiv preprint arXiv:2303.10420},
  year = {2023}
}

@misc{achiam2023gpt,
  title = {{GPT-4} Technical Report},
  author = {Achiam, Josh and Adler, Steven and Agarwal, Sandhini and Ahmad, Lama and Akkaya, Ilge and Aleman, Florencia Leoni and Almeida, Diogo and Altenschmidt, Janko and Altman, Sam and Anadkat, Shyamal and others},
  journal = {arXiv preprint arXiv:2303.08774},
  year = {2023}
}

@article{chowdhery2023palm,
  title = {{PaLM}: Scaling Language Modeling with Pathways},
  author = {Chowdhery, Aakanksha and Narang, Sharan and Devlin, Jacob and Bosma, Maarten and Mishra, Gaurav and Roberts, Adam and Barham, Paul and Chung, Hyung Won and Sutton, Charles and Gehrmann, Sebastian and others},
  journal = {Journal of Machine Learning Research},
  volume = {24},
  number = {240},
  pages = {1--113},
  year = {2023}
}

@misc{nijkamp2022codegen,
  title = {{CodeGen}: An Open Large Language Model for Code with Multi-Turn Program Synthesis},
  author = {Nijkamp, Erik and Pang, Bo and Hayashi, Hiroaki and Tu, Lifu and Wang, Huan and Zhou, Yingbo and Savarese, Silvio and Xiong, Caiming},
  journal = {arXiv preprint arXiv:2203.13474},
  year = {2022}
}

@misc{fried2022incoder,
  title = {{Incoder}: A Generative Model for Code Infilling and Synthesis},
  author = {Fried, Daniel and Aghajanyan, Armen and Lin, Jessy and Wang, Sida and Wallace, Eric and Shi, Freda and Zhong, Ruiqi and Yih, Wen-tau and Zettlemoyer, Luke and Lewis, Mike},
  journal = {arXiv preprint arXiv:2204.05999},
  year = {2022}
}

@misc{li2023starcoder,
  title = {{StarCoder}: May the Source Be with You!},
  author = {Li, Raymond and Allal, Loubna Ben and Zi, Yangtian and Muennighoff, Niklas and Kocetkov, Denis and Mou, Chenghao and Marone, Marc and Akiki, Christopher and Li, Jia and Chim, Jenny and others},
  journal = {arXiv preprint arXiv:2305.06161},
  year = {2023}
}

@misc{roziere2023code,
  title = {{Code Llama}: Open Foundation Models for Code},
  author = {Roziere, Baptiste and Gehring, Jonas and Gloeckle, Fabian and Sootla, Sten and Gat, Itai and Tan, Xiaoqing Ellen and Adi, Yossi and Liu, Jingyu and Sauvestre, Romain and Remez, Tal and others},
  journal = {arXiv preprint arXiv:2308.12950},
  year = {2023}
}

@inproceedings{liu2023verilogeval,
  title = {Verilogeval: Evaluating Large Language Models for {Verilog} Code Generation},
  author = {Liu, Mingjie and Pinckney, Nathaniel and Khailany, Brucek and Ren, Haoxing},
  booktitle = {2023 IEEE/ACM International Conference on Computer Aided Design ({ICCAD})},
  pages = {1--8},
  year = {2023},
  organization = {IEEE}
}

@misc{wang2024large,
  title = {Large Language Model for {Verilog} Generation with Golden Code Feedback},
  author = {Wang, Ning and Yao, Bingkun and Zhou, Jie and Wang, Xi and Jiang, Zhe and Guan, Nan},
  journal = {arXiv preprint arXiv:2407.18271},
  year = {2024}
}

@inproceedings{wong2024vgv,
  title = {{VGV}: {Verilog} Generation Using Visual Capabilities of Multi-Modal Large Language Models},
  author = {Wong, Sam-Zaak and Wan, Gwok-Waa and Liu, Dongping and Wang, Xi},
  booktitle = {2024 IEEE LLM Aided Design Workshop ({LAD})},
  pages = {1--5},
  year = {2024},
  organization = {IEEE}
}

@misc{nadimi2024multi,
  title = {A Multi-Expert Large Language Model Architecture for {Verilog} Code Generation},
  author = {Nadimi, Bardia and Zheng, Hao},
  journal = {arXiv preprint arXiv:2404.08029},
  year = {2024}
}

@misc{gao2024autovcoder,
  title = {{AutoVCoder}: A Systematic Framework for Automated {Verilog} Code Generation using LLMs},
  author = {Gao, Mingzhe and Zhao, Jieru and Lin, Zhe and Ding, Wenchao and Hou, Xiaofeng and Feng, Yu and Li, Chao and Guo, Minyi},
  journal = {arXiv preprint arXiv:2407.18333},
  year = {2024}
}

@inproceedings{thakur2023benchmarking,
  title = {Benchmarking Large Language Models for Automated {Verilog} {RTL} Code Generation},
  author = {Thakur, Shailja and Ahmad, Baleegh and Fan, Zhenxing and Pearce, Hammond and Tan, Benjamin and Karri, Ramesh and Dolan-Gavitt, Brendan and Garg, Siddharth},
  booktitle = {2023 Design, Automation \& Test in Europe Conference \& Exhibition ({DATE})},
  pages = {1--6},
  year = {2023},
  organization = {IEEE}
}

@inproceedings{lu2024rtllm,
  title = {{RTLLM}: An Open-Source Benchmark for Design {RTL} Generation with Large Language Models},
  author = {Lu, Yao and Liu, Shang and Zhang, Qijun and Xie, Zhiyao},
  booktitle = {2024 29th Asia and South Pacific Design Automation Conference ({ASP-DAC})},
  pages = {722--727},
  year = {2024},
  organization = {IEEE}
}

@misc{muennighoff2023octopack,
  title = {Octopack: Instruction Tuning Code Large Language Models},
  author = {Muennighoff, Niklas and Liu, Qian and Zebaze, Armel and Zheng, Qinkai and Hui, Binyuan and Zhuo, Terry Yue and Singh, Swayam and Tang, Xiangru and Von Werra, Leandro and Longpre, Shayne},
  journal = {arXiv preprint arXiv:2308.07124},
  year = {2023}
}

@article{liu2024your,
  title = {Is Your Code Generated by {ChatGPT} Really Correct? Rigorous Evaluation of Large Language Models for Code Generation},
  author = {Liu, Jiawei and Xia, Chunqiu Steven and Wang, Yuyao and Zhang, Lingming},
  journal = {Advances in Neural Information Processing Systems},
  volume = {36},
  year = {2024}
}

@inproceedings{lai2023ds,
  title = {{DS-1000}: A Natural and Reliable Benchmark for Data Science Code Generation},
  author = {Lai, Yuhang and Li, Chengxi and Wang, Yiming and Zhang, Tianyi and Zhong, Ruiqi and Zettlemoyer, Luke and Yih, Wen-tau and Fried, Daniel and Wang, Sida and Yu, Tao},
  booktitle = {International Conference on Machine Learning},
  pages = {18319--18345},
  year = {2023},
  organization = {PMLR}
}

@inproceedings{uniyal2024one,
  title = {One-to-Many Testing for Code Generation from (Just) Natural Language},
  author = {Uniyal, Mansi and Singh, Mukul and Verbruggen, Gust and Gulwani, Sumit and Le, Vu},
  booktitle = {Findings of the Association for Computational Linguistics: EMNLP 2024},
  pages = {15397--15402},
  year = {2024}
}

@misc{chen2021evaluating,
  title = {Evaluating Large Language Models Trained on Code},
  author = {Chen, Mark and Tworek, Jerry and Jun, Heewoo and Yuan, Qiming and Pinto, Henrique Ponde De Oliveira and Kaplan, Jared and Edwards, Harri and Burda, Yuri and Joseph, Nicholas and Brockman, Greg and others},
  journal = {arXiv preprint arXiv:2107.03374},
  year = {2021}
}

@misc{ren2020codebleu,
  title = {{CodeBLEU}: A Method for Automatic Evaluation of Code Synthesis},
  author = {Ren, Shuo and Guo, Daya and Lu, Shuai and Zhou, Long and Liu, Shujie and Tang, Duyu and Sundaresan, Neel and Zhou, Ming and Blanco, Ambrosio and Ma, Shuai},
  journal = {arXiv preprint arXiv:2009.10297},
  year = {2020}
}

@misc{abdelatty2024metrex,
  title = {{MetRex}: A Benchmark for {Verilog} Code Metric Reasoning Using LLMs},
  author = {Abdelatty, Manar and Ma, Jingxiao and Reda, Sherief},
  journal = {arXiv preprint arXiv:2411.03471},
  year = {2024}
}

@misc{yang2025haven,
  title = {HaVen: Hallucination-Mitigated LLM for {Verilog} Code Generation Aligned with {HDL} Engineers},
  author = {Yang, Yiyao and Teng, Fu and Liu, Pengju and Qi, Mengnan and Lv, Chenyang and Li, Ji and Zhang, Xuhong and He, Zhezhi},
  journal = {arXiv preprint arXiv:2501.04908},
  year = {2025}
}

@inproceedings{kashanaki2024hdleval,
  title = {{HDLEval} Benchmarking {LLMs} for Multiple {HDL}s},
  author = {Kashanaki, Farzaneh Rabiei and Zakharov, Mark and Renau, Jose},
  booktitle = {2024 IEEE {LLM} Aided Design Workshop ({LAD})},
  pages = {1--5},
  year = {2024},
  organization = {IEEE},
  publisher = {IEEE}
}

@inproceedings{batten2024pyhdl,
  title = {{PyHDL-Eval}: An {LLM} Evaluation Framework for Hardware Design Using Python-Embedded DSLs},
  author = {Batten, Christopher and Pinckney, Nathaniel and Liu, Mingjie and Ren, Haoxing and Khailany, Brucek},
  booktitle = {Proceedings of the 2024 ACM/IEEE International Symposium on Machine Learning for {CAD}},
  pages = {1--17},
  year = {2024},
  publisher = {IEEE}
}

@inproceedings{qiu2024autobench,
  title = {{Autobench}: Automatic Testbench Generation and Evaluation Using {LLMs} for {HDL} Design},
  author = {Qiu, Ruidi and Zhang, Grace Li and Drechsler, Rolf and Schlichtmann, Ulf and Li, Bing},
  booktitle = {Proceedings of the 2024 ACM/IEEE International Symposium on Machine Learning for {CAD}},
  pages = {1--10},
  year = {2024}
}

@misc{
wangenben,
title={GenBen:A Genarative Benchmark for {LLM}-Aided Design},
author={Gwok-Waa Wan and Wang yubo and SamZaak Wong and jingyi zhang and Mengnv Xing and Zhe jiang and Nan Guan and ying wang and Ning Xu and Qiang Xu and Xi Wang},
year={2025},
url={https://openreview.net/forum?id=gtVo4xcpFI}
}

@inproceedings{liu2024openllm,
    author = {Liu, Shang and Lu, Yao and Fang, Wenji and Li, Mengming and Xie, Zhiyao},
    title = {{OpenLLM-RTL}: Open Dataset and Benchmark for LLM-Aided Design {RTL} Generation},
    booktitle = {IEEE/ACM International Conference on Computer-Aided Design (ICCAD)},
    year = {2024}
}

@misc{vijayaraghavan2024vhdl,
  title = {{VHDL-Eval}: A Framework for Evaluating Large Language Models in {VHDL} Code Generation},
  author = {Vijayaraghavan, Prashanth and Shi, Luyao and Ambrogio, Stefano and Mackin, Charles and Nitsure, Apoorva and Beymer, David and Degan, Ehsan},
  journal = {arXiv preprint arXiv:2406.04379},
  year = {2024}
}

@misc{hendrycks2021measuring,
  title = {Measuring Mathematical Problem Solving with the Math Dataset},
  author = {Hendrycks, Dan and Burns, Collin and Kadavath, Saurav and Arora, Akul and Basart, Steven and Tang, Eric and Song, Dawn and Steinhardt, Jacob},
  journal = {arXiv preprint arXiv:2103.03874},
  year = {2021}
}

@misc{zheng2023survey,
  title = {A Survey of Large Language Models for Code: Evolution, Benchmarking, and Future Trends},
  author = {Zheng, Zibin and Ning, Kaiwen and Wang, Yanlin and Zhang, Jingwen and Zheng, Dewu and Ye, Mingxi and Chen, Jiachi},
  journal = {arXiv preprint arXiv:2311.10372},
  year = {2023}
}

@inproceedings{wang2023review,
  title = {A Review on Code Generation with {LLM}s: Application and Evaluation},
  author = {Wang, Jianxun and Chen, Yixiang},
  booktitle = {2023 IEEE International Conference on Medical Artificial Intelligence ({MedAI})},
  pages = {284--289},
  year = {2023},
  organization = {IEEE}
}

@misc{hdlbits2017,
  author = {Alvin Tan},
  title = {{HDLBits}: Digital Circuits Exercises},
  year = {2017},
  url = {https://hdlbits.01xz.net/wiki/Problem_sets},
  note = {Accessed: 2025}
}

@misc{hurst2024gpt,
  title = {{GPT}-4o System Card},
  author = {Hurst, Aaron and Lerer, Adam and Goucher, Adam P and Perelman, Adam and Ramesh, Aditya and Clark, Aidan and Ostrow, AJ and Welihinda, Akila and Hayes, Alan and Radford, Alec and others},
  journal = {arXiv preprint arXiv:2410.21276},
  year = {2024}
}

@misc{jaech2024openai,
  title = {{OpenAI} {O1} System Card},
  author = {Jaech, Aaron and Kalai, Adam and Lerer, Adam and Richardson, Adam and El-Kishky, Ahmed and Low, Aiden and Helyar, Alec and Madry, Aleksander and Beutel, Alex and Carney, Alex and others},
  journal = {arXiv preprint arXiv:2412.16720},
  year = {2024}
}

@misc{touvron2023llama,
  title = {{LLaMA}: Open and Efficient Foundation Language Models},
  author = {Touvron, Hugo and Lavril, Thibaut and Izacard, Gautier and Martinet, Xavier and Lachaux, Marie-Anne and Lacroix, Timoth{\'e}e and Rozi{\`e}re, Baptiste and Goyal, Naman and Hambro, Eric and Azhar, Faisal and others},
  journal = {arXiv preprint arXiv:2302.13971},
  year = {2023}
}

@misc{bai2023qwen,
  title = {{Qwen} Technical Report},
  author = {Bai, Jinze and Bai, Shuai and Chu, Yunfei and Cui, Zeyu and Dang, Kai and Deng, Xiaodong and Fan, Yang and Ge, Wenbin and Han, Yu and Huang, Fei and others},
  journal = {arXiv preprint arXiv:2309.16609},
  year = {2023}
}

@misc{yang2024qwen2,
  title = {{Qwen}2.5 Technical Report},
  author = {Yang, An and Yang, Baosong and Zhang, Beichen and Hui, Binyuan and Zheng, Bo and Yu, Bowen and Li, Chengyuan and Liu, Dayiheng and Huang, Fei and Wei, Haoran and others},
  journal = {arXiv preprint arXiv:2412.15115},
  year = {2024}
}

@misc{hui2024qwen2,
  title = {{Qwen}2.5-Coder Technical Report},
  author = {Hui, Binyuan and Yang, Jian and Cui, Zeyu and Yang, Jiaxi and Liu, Dayiheng and Zhang, Lei and Liu, Tianyu and Zhang, Jiajun and Yu, Bowen and Lu, Keming and others},
  journal = {arXiv preprint arXiv:2409.12186},
  year = {2024}
}

@misc{jiang2024survey,
      title={A Survey on Large Language Models for Code Generation}, 
      author={Juyong Jiang and Fan Wang and Jiasi Shen and Sungju Kim and Sunghun Kim},
      year={2024},
      eprint={2406.00515},
      archivePrefix={arXiv},
      primaryClass={cs.CL},
      url={https://arxiv.org/abs/2406.00515}, 
}

@misc{jiang2023mistral,
  title = {{Mistral 7B}},
  author = {Jiang, Albert Q and Sablayrolles, Alexandre and Mensch, Arthur and Bamford, Chris and Chaplot, Devendra Singh and Casas, Diego de las and Bressand, Florian and Lengyel, Gianna and Lample, Guillaume and Saulnier, Lucile and others},
  journal = {arXiv preprint arXiv:2310.06825},
  year = {2023}
}

\end{document}